\documentclass[preprintnumbers,amsmath,amssymb,amsfonts,superscriptaddress,nofootinbib]{revtex4}
\usepackage{graphicx}
\usepackage{amsmath}
\usepackage[utf8]{inputenc}
\usepackage{amssymb}
\usepackage{amsfonts}
\usepackage{amsbsy}
\usepackage{url}
\usepackage{enumerate}
\usepackage[justification=raggedright]{caption}
\usepackage{bbm}
\usepackage{color}

\newcommand{\be}{\begin{eqnarray}}
\newcommand{\ee}{\end{eqnarray}}
\newcommand{\p}{\partial}

\begin{document}

\preprint{YGHP-18-05,  KEK-TH-2041}

\title{Constraints on two Higgs doublet models from domain walls}

\author{Minoru~Eto}
\affiliation{Department of Physics, Yamagata University, Kojirakawa-machi 1-4-12, Yamagata, Yamagata 990-8560, Japan}
\author{Masafumi~Kurachi}
\affiliation{Research and Education Center for Natural Sciences, Keio University, 4-1-1 Hiyoshi, Yokohama, Kanagawa 223-8521, Japan}
\affiliation{Theory Center, High Energy Accelerator Research Organization (KEK), Tsukuba, Ibaraki 305-0801, Japan}
\author{Muneto~Nitta} 
\affiliation{Department of Physics, Keio University, 4-1-1 Hiyoshi, Kanagawa 223-8521, Japan}
\affiliation{Research and Education Center for Natural Sciences, Keio University, 4-1-1 Hiyoshi, Yokohama, Kanagawa 223-8521, Japan}

\begin{abstract}
We show that there is a constraint on the parameter space of two Higgs doublet models that comes from the existence of the stable vortex-domain wall systems. The constraint is quite universal in the sense that it depends on only two combinations of Lagrangian parameters and, at tree level, does not depend on how fermions couple to two Higgs fields. Numerical solutions of field configurations of domain wall-vortex system are obtained, which provide a basis for further quantitative study of cosmology which involve such topological objects.
\end{abstract}

\maketitle


\section{Introduction}
The discovery of the Higgs boson at the Large Hadron Collider (LHC) at CERN~\cite{Aad:2012tfa, Chatrchyan:2012xdj} proved that the Standard Model (SM) of the elementary particles is an appropriate low-energy effective description of our Universe. Now we are in a position to take a step forward and pursue solutions to problems that are left unanswered by the SM. Those include masses of neutrinos, baryon asymmetry of the Universe, the origin of the dark matter, etc. Currently the Higgs sector of the SM is the experimentally least constrained part. Therefore, it is tempting to extend Higgs sector to accommodate possibilities for solving above mentioned problems.

Two Higgs doublet model (2HDM)~\cite{Branco:2011iw}, in which the two Higgs doublet fields ($\Phi_1$, $\Phi_2$) are introduced instead of only one, is the most popular extension of the Higgs sector. Many studies have been done to solve problems which can not be addressed by the SM. Two Higgs doublet fields are also required when one considers supersymmetric extension of the SM. 2HDM has four additional scalar degree of freedom in addition to 125 GeV Higgs boson ($h$). Those are charged Higgs bosons ($H^\pm$), CP-even Higgs bosons ($H$) and CP-odd Higgs boson ($A$). These additional scalars can be directly produced at LHC, though there is no signal so far today, placing lower bounds on masses of those additional scalar bosons. Those lower bounds highly depend on parameter choices of 2HDM as well as how SM fermions couple to $\Phi_1$ and $\Phi_2$ since cross section of specific production and decay process crucially depends on those. The situation is similar for bounds obtained from other theoretical and phenomenological constraints~\cite{Kling:2016opi,globalfit, Kanemura:2014bqa}.
Therefore the constraints on parameters of 2HDM, that are determining masses of additional scalars, are also highly model dependent as well.

 The purpose of the study in this paper is to place more universal constraint which is less dependent on the details of model setups. We discuss possible existence of vortices (or cosmic strings)~\cite{Hindmarsh:1994re,Vilenkin:2000jqa} and domain walls (or kinks) in the 2HDM. When stable domain walls exist as  solutions of the theory, they must have been created in the early Universe through the Kibble-Zurek mechanism~\cite{Kibble:1976sj,Zurek:1985qw,Vachaspati-book}, resulting in the overclosure problem. We use this fact to place constraints on parameters of the 2HDM. Unlike the SM admitting only unstable electroweak $Z$-strings~\cite{Nambu:1977ag,Vachaspati:1992fi}, 2HDM admits topologically stable $Z$-strings~\cite{Dvali:1993sg,Dvali:1994qf}  in addition to unstable non-topological $Z$-strings \cite{Earnshaw:1993yu, Ivanov:2007de}. It also admits CP domain walls~\cite{Battye:2011jj, Thesis} and membranes~\cite{Bachas:1995ip, Riotto:1997dk}. 
 Topological strings are attatched by domain walls or membranes~\cite{Kibble:1982dd,Vilenkin:1982ks}, resembling axion strings \cite{Kawasaki:2013ae}, 
axial strings in dense QCD \cite{Eto:2013hoa} and topological $Z$-strings~\cite{Chatterjee:2018znk} in the Georgi-Machacek model~\cite{Georgi:1985nv}. When a string is attached by a single domain wall, such the domain wall can quantum mechanically decay by creating a hole bounded by the above mentioned string~\cite{Preskill:1992ck,Vilenkin:2000jqa}. We point out that whether stable domain walls exist or not depends on only two combinations of model parameters, therefore the cosmological requirement places constraint on these two parameters.

The paper is organized as follows. After we introduce the 2HDM Lagrangian in the next section, we discuss various types of domain walls and membranes that exist in 2HDMs in Sec.~\ref{sec:wall}. There, we systematically classify the parameter space of the 2HDMs to five cases, and shape of field configurations of domain walls and membranes will be presented for representative parameters in each case. Some of them correspond to known domain walls and membranes, while there is a new type of membrane which has substructure around the wall (Case IV). In Sec.~\ref{sec:vortex}, we discuss the vortex solution in 2HDMs in the case that the model has symmetry under the relative phase $U(1)$ rotation of two Higgs fields. Then in Sec.~\ref{sec:dw-st}, we introduce terms that explicitly breaks above mentioned $U(1)$ symmetry, and discuss the appearance of domain wall-vortex complex system. In Sec.~\ref{sec:cosmology}, cosmological impact of the existence of stable domain wall-vortex system is discussed, and constraint on the parameter space is obtained. Section~\ref{sec:summary} is dedicated to the conclusion and future prospects.

\section{The model}
We introduce two $SU(2)$ doublets, $\Phi_1$ and $\Phi_2$, both with the hypercharge $Y=1$. The Lagrangian which describes the electroweak and the Higgs sectors is written as
\be
{\cal L} = - \frac{1}{4}B_{\mu\nu}B^{\mu\nu} - \frac{1}{4}W_{\mu\nu}^aW^{a\mu\nu}  + \sum_{i=1,2}
\left(D_\mu \Phi_i^\dagger D^\mu\Phi_i\right) - V(\Phi_1, \Phi_2).
\ee
Here, $B_{\mu\nu}$ and $W^a_{\mu\nu}$ describe field strength tensors of the hypercharge and the weak gauge interactions with $\mu$ ($\nu$) and $a$ being Lorentz and weak iso-spin indices, respectively. $D_\mu$ represents the covariant derivative acting on the Higgs fields. The most general form of the potential, $V(\Phi_1, \Phi_2)$, which is consistent with the electroweak gauge invariance is expressed as follows:
\be
V(\Phi_1, \Phi_2) &=& 
m_{11}^2\Phi_1^\dagger\Phi_1 
+ m_{22}^2 \Phi_2^\dagger \Phi_2 
- \left(m_{12}^2 \Phi_1^\dagger \Phi_2 + {\rm h.c.}\right) 
+ \frac{\beta_1}{2}\left(\Phi_1^\dagger\Phi_1\right)^2
+ \frac{\beta_2}{2}\left(\Phi_2^\dagger\Phi_2\right)^2 \nonumber\\
&+& \beta_3\left(\Phi_1^\dagger\Phi_1\right)\left(\Phi_2^\dagger\Phi_2\right) 
+ \beta_4 \left(\Phi_1^\dagger\Phi_2\right)\left(\Phi_2^\dagger\Phi_1\right) +\left\{\frac{\beta_5}{2}\left(\Phi_1^\dagger\Phi_2\right)^2 + {\rm h.c.}\right\}
\nonumber\\
&+&\left\{ \left[ \beta_6 \Phi_1^\dagger\Phi_1 + \beta_7 \Phi_2^\dagger\Phi_2 \right]\left(\Phi_1^\dagger\Phi_2\right) + {\rm h.c.}\right\}.
\label{eq:potential}
\ee
In order to keep Higgs-mediated flavor-changing neutral current processes under control, we impose a (softly-broken) ${\mathbb Z}_2$ symmetry,
$\Phi_1 \to +\Phi_1$ and $\Phi_2 \to -\Phi_2$, on the potential~\cite{Glashow:1976nt}. This is achieved by taking $\beta_6 = \beta_7 = 0$ in Eq.~(\ref{eq:potential}). 
We also assume, as is often done in the literature, that the Lagrangian is invariant under the following CP transformation:
\be
\Phi_i \to i \sigma_2 \Phi_i^\ast.
\label{eq:CP}
\ee
CP invariant Lagrangian is obtained by taking $m_{12}^2$ and $\beta_5$ to be real. \footnote{Since the fermion sector breaks the CP symmetry, even if we take $m_{12}^2$ and $\beta_5$ to be real values, fermion loop contributions renormalize these parameters to be complex numbers. We will discuss the effect of explicit CP breaking terms in Sec.~\ref{sec:cosmology}.} We should note here that the sign of $m_{12}^2$ is irrelevant to physics since it can always be changed by a field redefinition such as $\Phi_2 \to - \Phi_2$. In this letter, we define Higgs fields in such a way that $m_{12}^2$ becomes non-negative.
Now, we consider the situation that both Higgs fields develop vacuum expectation values (VEVs) as~\footnote{In the literature, it is often expressed as $ \Phi_1 =  e^{-i\zeta}(0, v_1)^{T}, \Phi_2 = (0, v_2)^{T}$, or $ \Phi_1 =  (0, v_1)^{T}, \Phi_2 = e^{i\zeta}(0, v_2)^{T}$. These are equivalent to Eq.~(\ref{eq:VEV}) with $\alpha=\zeta/2$ up to gauge transformation.
}
\be
\Phi_1 =  e^{-i\alpha}\left(\begin{array}{c} 0 \\ v_1\end{array}\right),\quad
\Phi_2 =  e^{i\alpha}\left(\begin{array}{c} 0 \\ v_2\end{array}\right).
\label{eq:VEV}
\ee
Then the electroweak scale, $v_{\rm EW}$ ($\simeq $ 246 GeV), can be expressed by these VEVs as $v_{\rm EW}^2/2 = (v_1^2 + v_2^2)$. When  $\alpha$, the relative $U(1)$ phase of two Higgs fields, takes 0 (mod $\pi/2$), the vacuum is invariant under the CP transformation in Eq.~(\ref{eq:CP}) (up to the field redefinition $\Phi_i \to -\Phi_i$). When $\alpha$ takes a value other than $0$ mod $\pi/2$, the vacuum is not invariant under the CP transformation, therefore the CP symmetry is spontaneously broken. It is maximally broken when $\alpha$ becomes $\pi/4$ mod $\pi/2$.

For later use, we introduce a notation that two Higgs fields are combined into a single two-by-two matrix form, $H$, defined as
\be
H = \left( i\sigma_2 \Phi_1^*,\ \Phi_2\right).
\ee
The field $H$ transforms under the electroweak $SU(2)_L \times U(1)_Y$ symmetry as 
\be
H  &\to&  e^{i\alpha_a(x) \frac{\sigma_a}{2}} H e^{-i \beta(x) \frac{\sigma_3}{2}},
\ee
therefore the covariant derivative of $H$ is expressed as
\be
D_\mu H &=& \p_\mu H - g\frac{i}{2}\sigma_a W_\mu^a H + g'\frac{i}{2}H\sigma_3B_\mu.
\ee
The VEV of $H$ is expressed by a diagonal matrix $\langle H \rangle = e^{i\alpha}{\rm diag}(v_1,v_2)$, and the potential can be written by using $H$ as follows:
\be
&& \hspace{-0.5cm}V
= \frac{m_{11}^2 + m_{22}^2}{2} {\rm Tr}\left(H^\dagger H\right) 
- \frac{m_{11}^2 - m_{22}^2}{2} {\rm Tr}\left(H^\dagger H \sigma_3\right) -  m_{12}^2 \left( \det H + {\rm h.c.}\right)\nonumber\\
&&+ \frac{2(\beta_1+\beta_2)+3\beta_3}{12}{\rm Tr}\left(H^\dagger H H^\dagger H\right) + \frac{2(\beta_1+\beta_2)-3\beta_3}{12}{\rm Tr}\left(H^\dagger H\sigma_3 H^\dagger H\sigma_3\right)\nonumber\\
&&- \frac{\beta_1 - \beta_2}{3} {\rm Tr}\left(H^\dagger H\sigma_3 H^\dagger H\right) 
+ \left(\beta_3 + \beta_4\right) \det\left(H^\dagger H \right) 
+ \left(\frac{\beta_5}{2}\det H^2 + {\rm h.c.}\right).
\ee
\section{Domain walls and membranes}
\label{sec:wall}
\begin{figure}[t]
\begin{center}
\includegraphics[width=0.4\linewidth]{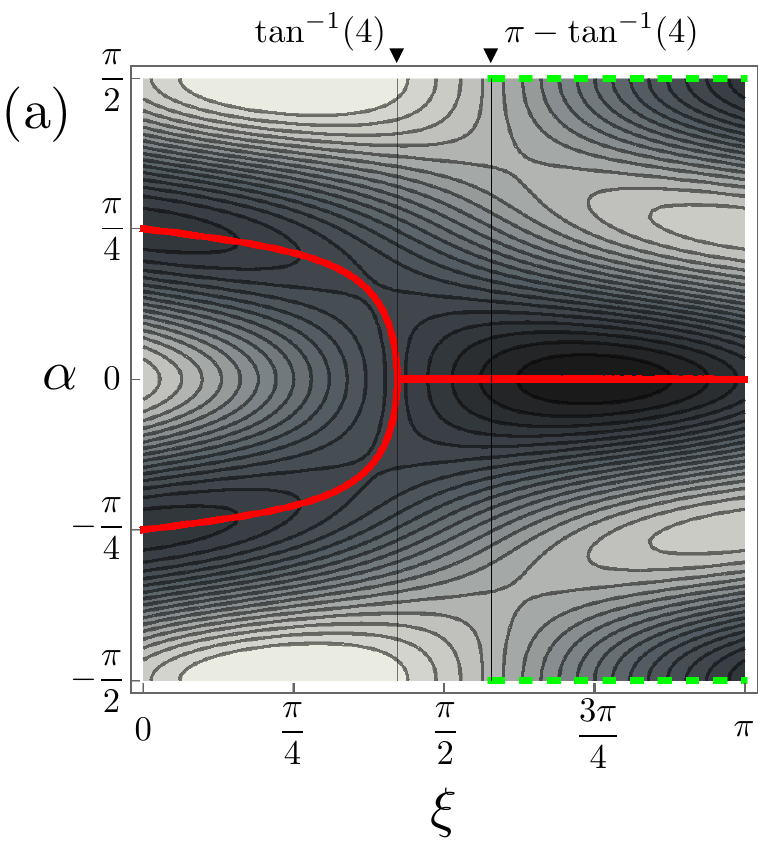}\vspace{5mm}\\
\includegraphics[width=\linewidth]{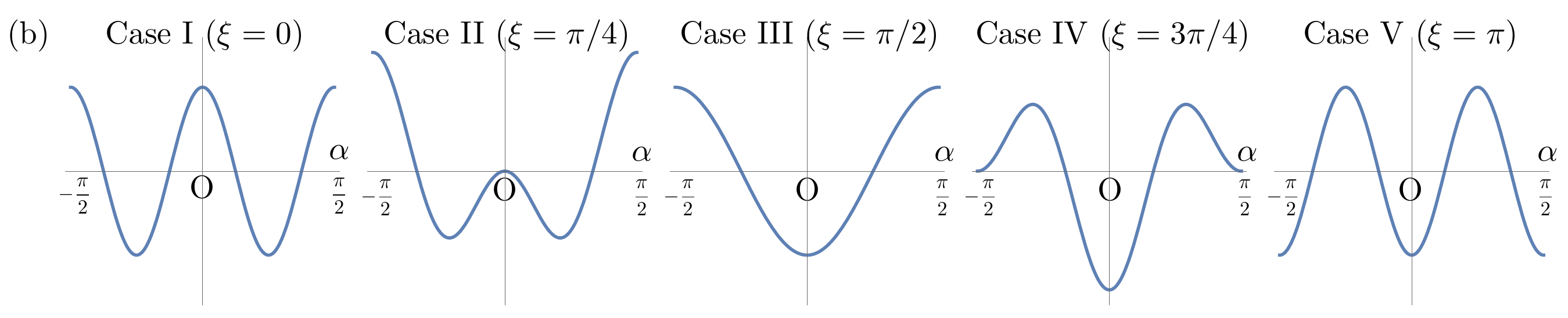}
\caption{(a) Contour plot of  the $U(1)_a$-dependent part of the potential. The horizontal axis is $\xi \in [0,\pi]$ while the vertical axis is $\alpha \in [-\pi/2,\pi/2]$. The darker the color is, the lower the potential is. Bottom of the potential and the local minimum are indicated by thick-solid and thick-dashed lines, respectively.
(b) Typical slices of $V_\xi(\alpha)$ for $\alpha \in [-\pi/2,\pi/2]$ with $\xi=0, \pi/4, \pi/2, 3\pi/4$ and $\xi=\pi$.
}
\label{fig:det_det2_parameter}
\end{center}
\end{figure}

In this section, we discuss the relative $U(1)$ phase dependence of the potential and the existence of wall solutions associated with non-trivial winding of the phase. To see this, let us substitute the Higgs fields which have forms shown in Eq.~(\ref{eq:VEV}) into the relative phase dependent terms in the potential, namely terms proportional to $m_{12}^2$ and $\beta_5$:
\be
V_{\xi}(\alpha) &=& -2 m_{12}^2 v_1 v_2 \cos2\alpha + \beta_5 v_1^2v_2^2\cos 4\alpha \nonumber\\
&=& (v_1v_2)^2 \sqrt{4\left(m_{12}^2/v_1 v_2\right)^2+\beta_5^2} \nonumber\\
 &&  \quad\quad  \left(-\sin\xi\cos2\alpha + \cos\xi\cos4\alpha\right).
\label{eq:potentialU1a}
\ee
Here, we have defined the angle $\xi$ as follows:
\be
\sin\xi &=&  \frac{2 (m_{12}^2/v_1 v_2)}{\sqrt{4(m_{12}^2/v_1v_2)^2+\beta_5^2}}, \nonumber\\
\cos\xi &=& \frac{\beta_5}{\sqrt{4(m_{12}^2/v_1v_2)^2+\beta_5^2}}. 
\ee
This is a one parameter family of the double sine-Gordon potential. In Fig.~\ref{fig:det_det2_parameter}~(a), we show the phase dependent part of the potential (the last line in Eq.~(\ref{eq:potentialU1a})).
In the figure, the darker the color is, the lower the potential is, and the bottom (vacua) of the potential is indicated by thick-solid curves. The local minima  (false vacua) are also indicated by thick-dashed lines in the figure. The potential has at most two local maxima. Note that since a field with $\alpha = \delta$ (where $\delta$ is an arbitrary real value) and that with $\alpha = \delta + \pi$ are physically equivalent up to gauge transformation ($\pi$ rotation in $\sigma_3$ component of $SU(2)_L$ gauge transformation), the potential has a periodicity of $\pi$ in the direction of $\alpha$. 
Therefore, we show the potential in the range of $-\pi/2 \le \alpha \le \pi/2$. Since we defined two Higgs fields in a way that $m_{12}^2$ becomes non-negative, $\xi$ takes its value in the range of $0 \le \xi \le \pi$. 
In Fig.~\ref{fig:det_det2_parameter}~(b), we show typical slices of $V_\xi(\alpha)$ for $\alpha \in [-\pi/2,\pi/2]$ with $\xi=0, \pi/4, \pi/2, 3\pi/4$ and $\xi=\pi$.
Depending on the relative importance of $m_{12}^2$ term and $\beta_5$ term, which is characterized by the angle $\xi$, the vacuum structure can be classified into the following five cases:\\

\noindent$\bullet$ Case I: $\xi =0$

\begin{figure}[t]
\begin{center}
\includegraphics[width=\linewidth]{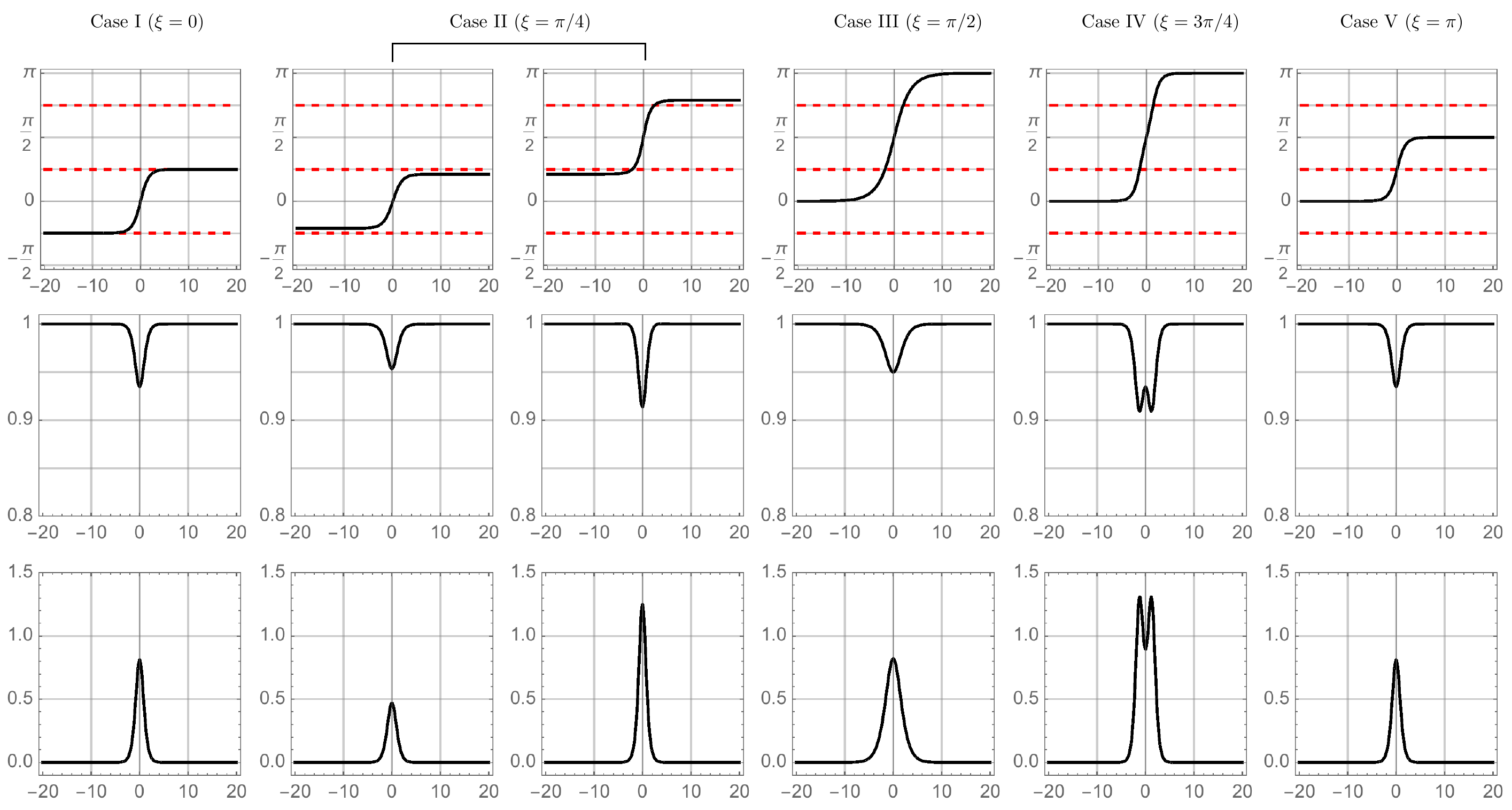}
\caption{Domain wall/membrane configurations (top two rows) and the energy distribution (the bottom row) for $\xi=0, \pi/4, \pi/2, 3\pi/4$ and $\xi=\pi$. The top row shows coordinate dependence of relative $U(1)$ phase, while the second row shows magnitude of the radial component, namely $\sqrt{h_1^2+h_2^2}/\sqrt{v_1^2+v_2^2}$, where $h_1$ and $h_2$ are the real part of neutral component of two Higgs fields. All the quantities that have mass dimension are normalized by $\sqrt{-m_{11}^2}/2$.
For all the figures, we commonly took $m_{11}^2 = m_{22}^2 = -4$, $\beta_1 = \beta_2 = 3\beta_3/4 =
9/2$, and $\beta_4=-8$, and 
horizontal axis represents coordinate which is orthogonal to the wall. 
The parameters $m_{12}^2$ and $\beta_5$ are  chosen as 
$(m_{12}^2/(-m_{11}^2/4),\beta_5) = (0,1/5),\, (1/10,1/5),\,(1/10,0),\,(1/10,-1/5),\ (0,-1/5)$ for
$\xi = 0,\pi/4, \pi/2, 3\pi/4, \pi$, respectively.
In figures in top row, $\alpha=\pm \pi/4$ (and $\alpha=3\pi/4$, which is equivalent to $\alpha=-\pi/4$ up to gauge transformation) are indicated by dashed lines, that correspond to the relative phases that maximally break the CP symmetry.
}
\label{fig:dw}
\end{center}
\end{figure}

There are two degenerate vacua at $\alpha = \pm \pi/4$. (See the first panel of Fig.~\ref{fig:det_det2_parameter}~(b).) The vacuum with $\alpha=-\pi/4$ is obtained by applying the CP transformation to the vacuum with $\alpha=\pi/4$, and vice versa. Thus, the CP symmetry is spontaneously broken in this case. Since the CP symmetry is a discrete symmetry, and it is broken spontaneously, we expect that there exists domain wall configurations that connect these two different vacua. In the literature, this is often called a CP domain wall~\cite{Battye:2011jj, Thesis}. Three panels in the first column in Fig.~\ref{fig:dw} show an example of domain wall solution. Since the potential has sine-type periodicity in terms of the relative phase, the domain wall solution is similar to that of the sine-Gordon kink. General feature of the CP domain wall is that the CP symmetry is restored on top of the wall.\\

\noindent$\bullet$ Case II: $0 < \xi< \tan^{-1}(4)$

There are two degenerate vacua at $\alpha = \pm \delta$, where $\delta$ takes a value in the range of $0 < \delta < \pi/4$. (See the second panel of Fig.~\ref{fig:det_det2_parameter}~(b) as a typical example of this case.) Similar to the Case I, the CP symmetry is spontaneously broken, therefore the CP domain wall appears in this case as well.  In the second and the third columns  in Fig.~\ref{fig:dw}, we show two types of CP domain wall for the case of $\xi=\pi/4$. The reason why there are two types of solutions is because the hight of two energy barriers (at $\alpha=0$ and $\pi/2$) are different. The second column of Fig.~\ref{fig:dw} shows the domain wall that crosses lower barrier, while the third column shows that crosses higher barrier. \\

\noindent$\bullet$ Case III: $\tan^{-1}(4) \le \xi< \pi - \tan^{-1}(4)$

There is only one global minimum at $\alpha=0$. (See the third panel of Fig.~\ref{fig:det_det2_parameter}~(b) as a typical example of this case.) Since there is no discrete symmetry which is spontaneously broken, there is no domain wall configuration in this region. However, the periodicity of the potential in terms of $\alpha$ provides a sine-Gordon kink-type configuration, or a membrane~\cite{Bachas:1995ip}, that interpolates $\alpha=0$ at, say, $x=-\infty$ to $\alpha=\pi$  at $x=\infty$. See the fourth column in Fig.~\ref{fig:dw} for this type of configuration in the case of $\xi=\pi/2$.  The qualitative difference compared to the domain wall discussed in Cases I and II is that the membrane configuration connects physically the same configuration. (Note that, as we mentioned above, the field with $\alpha=\pi$ is equivalent to that with $\alpha=0$ up to the gauge transformation.) Therefore it can decay to trivial vacuum configuration through quantum tunneling. How it can happen will be discussed later. Another interesting feature of the membrane is that the CP symmetry is locally broken near the membrane. The cosmological impact of the existence of such a local source of CP violation was discussed in Ref.~\cite{Riotto:1997dk} in the context of 2HDM in the Minimal Supersymmetric Standard Model.\\

\noindent$\bullet$ Case IV: $\pi - \tan^{-1}(4) \le \xi<\pi$

In addition to a global minimum at $\alpha=0$, there is a local (metastable) minimum at $\alpha = \pi/2$ (or at $\alpha = -\pi/2$). (See the forth panel of Fig.~\ref{fig:det_det2_parameter}~(b) as a typical example of this case.) Situation is similar to the Case III in a sense that there exist membrane configurations. The difference is that the membrane has a substructure due to the metastable vacua. This type of membrane has not been discussed in the literature. In the fifth column of Fig.~\ref{fig:dw}, we show the shape of configuration and energy distribution for this type of membrane solution in the case of $\xi=3\pi/4$. \\
 
\noindent$\bullet$ Case V: $\xi = \pi$

Similar to the Cases I and II, there are two degenerate vacua in this case (See the fifth panel of Fig.~\ref{fig:det_det2_parameter}~(b).): one at $\alpha=0$ and the other at $\alpha = \pi/2$ (or, equivalently, at $\alpha=-\pi/2$). The vacuum field with $\alpha=\pi/2$ is obtained by applying the transformation $\Phi_1 \to -i\Phi_1$, $\Phi_2 \to i\Phi_2$ to that with $\alpha=0$, and vice versa. This transformation is, up to gauge transformation ($\pi/2$ rotation in $\sigma_3$ component of $SU(2)_L$ gauge transformation), equivalent to a familiar form of ${\mathbb Z}_2$ transfomation: $\Phi_1 \to \Phi_1$, $\Phi_2 \to -\Phi_2$. Since this ${\mathbb Z}_2$ symmetry is spontaneously broken, we expect the appearance of ${\mathbb Z}_2$ domain wall which divides domains of $\alpha=0$ and $\alpha=\pi/2$. See the sixth column in Fig.~\ref{fig:dw} for such domain wall configuration. It should be noted that the shape of the domain wall solution in this case is same as that of the Case I, however, solutions in these two cases are qualitatively different from the perspective of the CP symmetry: unlike the solution in the Case I, in this case, the CP symmetry is maximally broken on the top of the domain wall, while the configuration approaches to the CP symmetric field value away from the wall.\\

\begin{figure}[t]
\begin{center}
\includegraphics[width=0.8\linewidth]{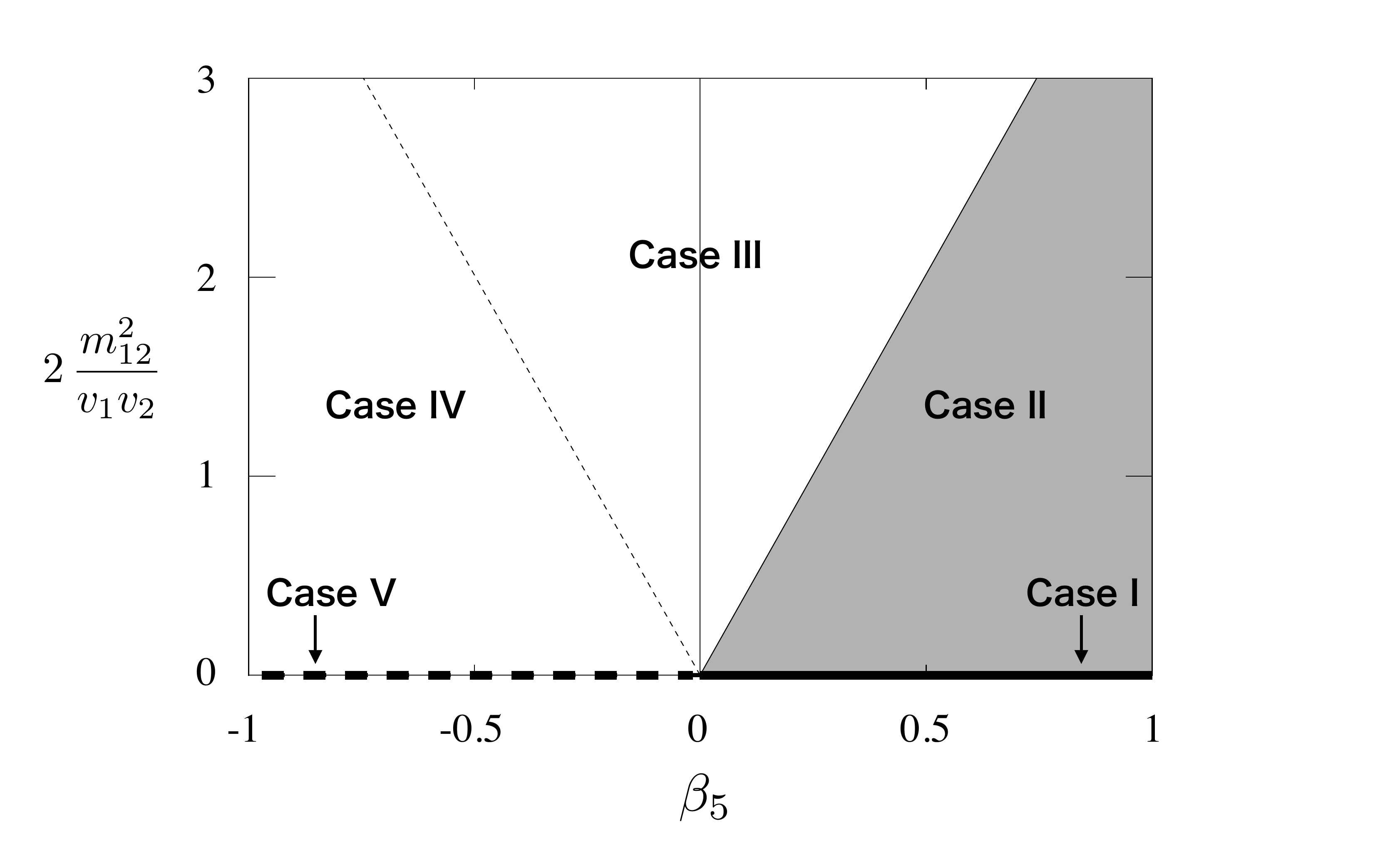}
\caption{Classification of the parameter space in terms of the vacuum structure. 
Shaded area (indicated by ``Case II") and the thick-solid line (indicated by ``Case I") and thick-dashed line (indicated by ``Case V") are cosmologically excluded due to the existence of stable domain walls.}
\label{fig:constraint}
\end{center}
\end{figure}

The regions on the $\beta_5$ - $(2m_{12}^2/v_1v_2)$ parameter space that correspond to each case are shown in Fig.~\ref{fig:constraint}. The shaded area in the figure corresponds to the Case II. The unshaded area is divided into two regions: the right side of the thin dashed line corresponds to the Case III, while left to the thin dashed line corresponds to Case IV. Case I is on the positive $\beta_5$ axis indicated by thick solid line, while Case V is on the negative $\beta_5$ axis indicated by thick dashed line.
The stability of domain walls/membranes, as well as cosmological impact in each case will be discussed later.

\section{Vortices and Z-strings}
\label{sec:vortex}

Now we discuss another type of topological object, vortices, which exist in 2HDMs. To explain it, we first consider the situation that terms proportional to $m_{12}^2$ and $\beta_5$ are absent. In this situation, the potential has an additional $U(1)_a$ symmetry which rotates the relative phase of $\Phi_1$ and $\Phi_2$ as
\be
U(1)_a:\ H \to e^{i\delta} H.
\ee
We note that this symmetry is spontaneously broken in the vacuum, and  corresponding Nambu-Goldstone (NG) mode appears. Actually, in the limit of $m_{12}^2 = \beta_5 = 0$, the mass of the CP-odd Higgs, $m_A$, vanishes, suggesting that the CP-odd Higgs field is identified with this NG mode.

In Ref.~\cite{Dvali:1993sg} it was pointed out that thanks to this extra $U(1)_a$ symmetry, topologically stable vortices, or cosmic strings, with $Z$ fluxes can exist in 2HDMs. The topological stability is guaranteed by the nontrivial homotopy $\pi_1(U(1)_a) = \mathbb{Z}$.
To describe the $Z$ strings, let us take $W_\mu^{\pm} = A_\mu = 0$, 
where we have defined 
$A_\mu = B_\mu \cos\theta_W + W_\mu^3 \sin\theta_W$ and 
$Z_\mu = W^3_\mu \cos\theta_W - B_\mu \sin\theta_W$ with
$\tan\theta_W = g'/g$.
Then the covariant derivative reads
\be
D_\mu H = \left(\p_\mu - \frac{g}{\cos\theta_W}\frac{i\sigma_3}{2}Z_\mu\right)H.
\ee
Now, it is easy to observe two different kinds of $Z$ strings 
for a generic choice of the parameters (with $m_{12}^2 = \beta_5=0$). 
The one which we call the $(1,0)$-string parallel to the $x^3$ axis is given by
\be
H^{(1,0)} &=& \left(
\begin{array}{cc}
v_1 f_1^{(1,0)}(r)e^{i\theta} & 0\\
0 & v_2 f_2^{(1,0)}(r)
\end{array}
\right),
\label{eq:10vortexH}\\
Z_i^{(1,0)} &=& \frac{-2v_1^2}{v_1^2 + v_2^2}
\frac{\cos\theta_W}{g}\epsilon_{ij}\frac{x^j}{r^2}\left(1-w^{(1,0)}(r)\right),
\label{eq:10vortexZ}
\ee
with $x^1 + i x^2 = r e^{i\theta}$.
The other which we call the $(0,1)$-string is given as
\be
H^{(0,1)} &=& \left(
\begin{array}{cc}
v_1 f_1^{(0,1)}(r) & 0\\
0 & v_2 f_2^{(0,1)}(r) e^{i\theta}
\end{array}
\right),\\ 
Z_i^{(0,1)} &=& \frac{2v_2^2}{v_1^2 + v_2^2} 
\frac{\cos\theta_W}{g}\epsilon_{ij}\frac{x^j}{r^2}\left(1-w^{(0,1)}(r)\right).
\ee
The boundary conditions imposed on profile functions are 
$f_1^{(1,0)}(0)
=0$, $f_2'^{(1,0)}(0)
=0$, $w^{(1,0)}(0)
=1$, $f_1^{(1,0)}(\infty)=f_2^{(1,0)}(\infty)
=1$, $w^{(1,0)}(\infty)
=0$. 
Similar conditions are required for the $(0,1)$ string.
Since vortices wind around the $U(1)_a$ symmetry which is a global symmetry, 
they are global cosmic strings having logarithmically divergent tension in infinite space.
Here, the ansatz for the $Z_i$ is determined in such a way that the logarithmic divergence
from the kinetic term of the Higgs $\int d^2x\ {\rm Tr}(D_iH D_iH^\dagger)$ 
is minimized. Then, the logarithmic divergence of the string tension is the same 
for both the (1,0) and (0,1) strings as
\be
T_Z = 2\pi \frac{v_1^2v_2^2}{v_1^2+v_2^2} \log \Lambda+ \cdots,
\label{eq:logdiv}
\ee
where $\Lambda$ is an IR cutoff and the ellipses stand for finite parts. 
Nature as the global string can directly be
seen in the asymptotic behavior, say, $H^{(1,0)}$ at $r\to\infty$:
\be
H^{(1,0)} \to 
\left(
\begin{array}{cc}
v_1 e^{i\theta} & 0\\
0 & v_2 
\end{array}
\right)
= e^{\frac{i}{2}\theta} e^{\frac{i\sigma_3}{2}\theta}
\left(
\begin{array}{cc}
v_1 & 0\\
0 & v_2 
\end{array}
\right).
\label{eq:linear}
\ee
Thus, we find the global $U(1)_a$ phase depends on $\theta$ as $\alpha = \frac{\theta}{2}$.
Therefore, the $U(1)_a$ phase rotates but only by $\pi$ when we go once around the $(1,0)$ string.
On the other hand, the $Z$ flux condensed in the string is finite and given by
\be
\Phi_Z^{(1,0)} &=&  \int d^2x\, Z_{12}^{(1,0)} = \frac{2\pi \cos\theta_W}{g} \frac{2 v_1^2}{v_1^2+v_2^2},\\
\Phi_Z^{(0,1)} &=&  \int d^2x\, Z_{12}^{(0,1)} = - \frac{2\pi \cos\theta_W}{g} \frac{2 v_2^2}{v_1^2+v_2^2}.
\ee
Note that $\Phi_Z^{(1,0)} - \Phi_Z^{(0,1)} = 4\pi \cos\theta_W/g $ 
gives a unit quantized flux of a non-topological $Z$ string while 
each carries a fractionally quantized flux. Full numerical solutions are presented in Ref.~\cite{Eto:2018tnk} in which physical properties are also discussed in detail.\\

\section{Domain wall-vortex system}
\label{sec:dw-st}

\begin{figure}[t]
\begin{center}
\includegraphics[width=\linewidth]{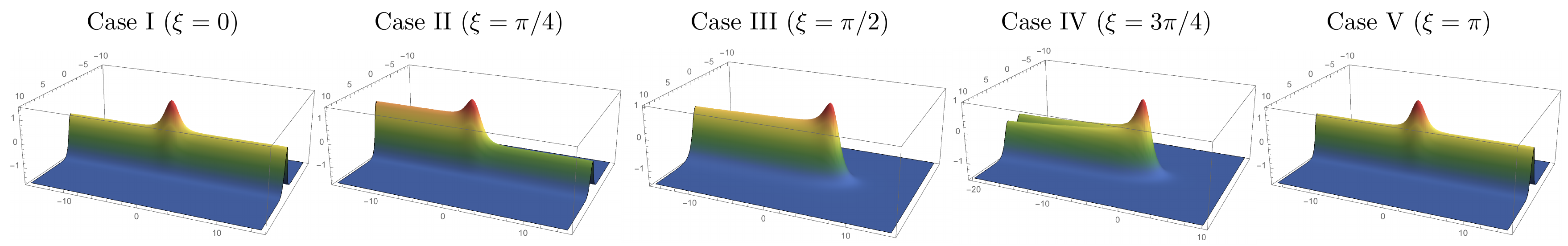}
\caption{Energy distribution for wall-string systems (on the plane that is orthogonal to the string) for $\xi=0, \pi/4, \pi/2, 3\pi/2, \pi$. All the quantities that have mass dimension are normalized by $\sqrt{-m_{11}^2}$.
For all the figures, we commonly took $m_{11}^2 = m_{22}^2 = -1$, $\beta_1 = \beta_2 = 3\beta_3/8 =
9/2$, and $\beta_4=-2$.
The parameters $m_{12}^2$ and $\beta_5$ are  chosen as 
$(m_{12}^2/(-m_{11}^2),\beta_5) = (0,2/5),\, (1/5,2/5),\,(1/5,0),\,(1/10,-1/5),\ (0,-2/5)$ for
$\xi = 0,\pi/4, \pi/2, 3\pi/4, \pi$, respectively.
}
\label{fig:wall_vortex}
\end{center}
\end{figure}

In the previous section, we have discussed vortex configurations in 2HDMs in a specific situation where $U(1)_a$ symmetry is exact, in which case the CP-odd Higgs becomes the NG boson. In order to make the model realistic, we need to give a mass to the CP-odd Higgs boson. This is done by switching on explicit $U(1)_a$ breaking terms, namely terms that are proportional to $m_{12}^2$ and $\beta_5$.
Here, we discuss the effect of these explicit $U(1)_a$ breaking terms on the vacuum structure, and show that the domain walls are created around the vortex~\footnote{The formation of domain walls around the vortex in 2HDM was first discussed in Ref.~\cite{Dvali:1993sg}.}.
Now let us consider how the vortex configuration should behave under the existence of this potential. For this purpose, let us consider the $(1, 0)$-string solution shown in Eqs.~(\ref{eq:10vortexH}) and (\ref{eq:10vortexZ}) as an example. (The discussion below can be applied for the case of $(0, 1)$-vortex in a similar way.) Let us look at the asymptotic behavior of $H^{(1,0)}$:
When we switch on the $U(1)_a$ breaking terms, we expect that the phase appearing in the upper-left element of the vortex field configuration depends on the angle coordinate $\theta$ non-trivially, rather than a simple linear dependence as Eq.~(\ref{eq:linear}). 
We denote this $\theta$ dependence by a function $\hat{\theta}(\theta)$:
\be
\hat{H}^{(1,0)}(r=\infty) 
= e^{\frac{i}{2}\hat{\theta}(\theta) }\ 
e^{\frac{i\sigma_3 }{2}\hat{\theta}(\theta)}
\left(
\begin{array}{cc}
v_1 & 0\\
0 & v_2 
\label{eq:hatH}
\end{array}
\right)
\ee
We expressed the phase function by the product of two rotations, namely $U(1)_a$ global rotation 
\be
\alpha = \frac{\hat \theta(\theta)}{2},
\label{eq:alpha}
\ee
and 
a $U(1)_Y$ gauge rotation. The latter is less important for domain walls (but important for flux and single-valuedness) since the potential is invariant under this rotation, while the former plays important role since the potential now depends on the $U(1)_a$ phase as shown in Eq.~(\ref{eq:potentialU1a}). When we consider the field configuration on  a closed loop around the vortex which is parameterized by $\theta$ in the range of $0 \leq \theta \leq 2\pi$, the single-valuedness of the field requires that $\hat{\theta}(\theta)$ is a continuous function which takes values $\hat{\theta}(\theta=0) = \delta$ and $\hat{\theta}(\theta=2\pi) = \delta + 2\pi$. Here, $\delta$ is an arbitrary constant, which is irrelevant for the discussion here thanks to the periodicity of the potential in Eq.~(\ref{eq:potentialU1a}). Therefore we take $\delta=-\pi$ in the following discussion. Then the $U(1)_a$ phase in Eq.~(\ref{eq:alpha}) takes its value from $-\pi/2$ to $\pi/2$ along the closed loop. Therefore the shape of potential in Eq.~(\ref{eq:potentialU1a}) in the range of $-\pi/2 \le \alpha \le \pi/2$ controls the form of the field configuration around the vortex. 
Passing a local maximum costs extra energy, which implies that a domain wall bounded by the string appears.
Now we divide $\xi$ into two classes of regions as follows:
\be
{\rm Region\ I\ (Cases\  I,\  II\  and\  V):} && 0 \le \xi < \tan^{-1}\left(4\right) \nonumber\\
&& {\rm and}\ \ \xi = \pi  , \nonumber\\
{\rm Region\ II\ (Cases\  III\  and\  IV):} && \tan^{-1}\left(4\right) \le \xi < \pi.\nonumber
\ee 
These are classified by the number of global minimum. 
When $\xi$ takes a value in Region~I (see the first, the second and the fifth panels of Fig.~\ref{fig:det_det2_parameter}~(b) for typical shapes of the potential in Region~I), 
 $V_\xi(\alpha)$ has two degenerate discrete vacua in the range of $-\pi/2 \le \alpha \le \pi/2$. This indicates the existence of two domain walls attached to the vortex. Since the height of two local maxima
are in general different, the tensions of two domain walls are imbalanced. (See the second panel of Fig.~\ref{fig:det_det2_parameter}~(b), where the energy distribution of the system for the case of $\xi=\pi/4$ is shown.) Therefore, the string is pulled toward the large tension domain wall.
The string can be static only for $\xi = 0$ and $\pi$ where the two domain walls 
have the same tension. (See the first and the fifth panels of Fig.~\ref{fig:det_det2_parameter}~(b).) 
On the other hand, when $\xi$ takes a value in Region~II, the potential has only one global minimum. In this case, only one domain wall appears that attaches to the vortex. (See the third and the fourth panels of Fig.~\ref{fig:det_det2_parameter}~(b) for the cases of $\xi=\pi/2$ and $3\pi/4$.) Precisely speaking, when there exists a false vacuum,  the domain wall has a substructure that it consists of two constituent domain walls. In Fig.~\ref{fig:wall_vortex}, energy distributions for wall-string systems (on the plane that is orthogonal to the string) for $\xi=0, \pi/4, \pi/2, 3\pi/2, \pi$ are shown. Qualitative difference of domain wall-vortex system in two classes of parameter regions makes a significant difference in constraining the model.
\section{Cosmological constraint on the $U(1)_a$ breaking parameters}
\label{sec:cosmology}
\begin{figure}[t]
\begin{center}
\includegraphics[width=0.7\linewidth]{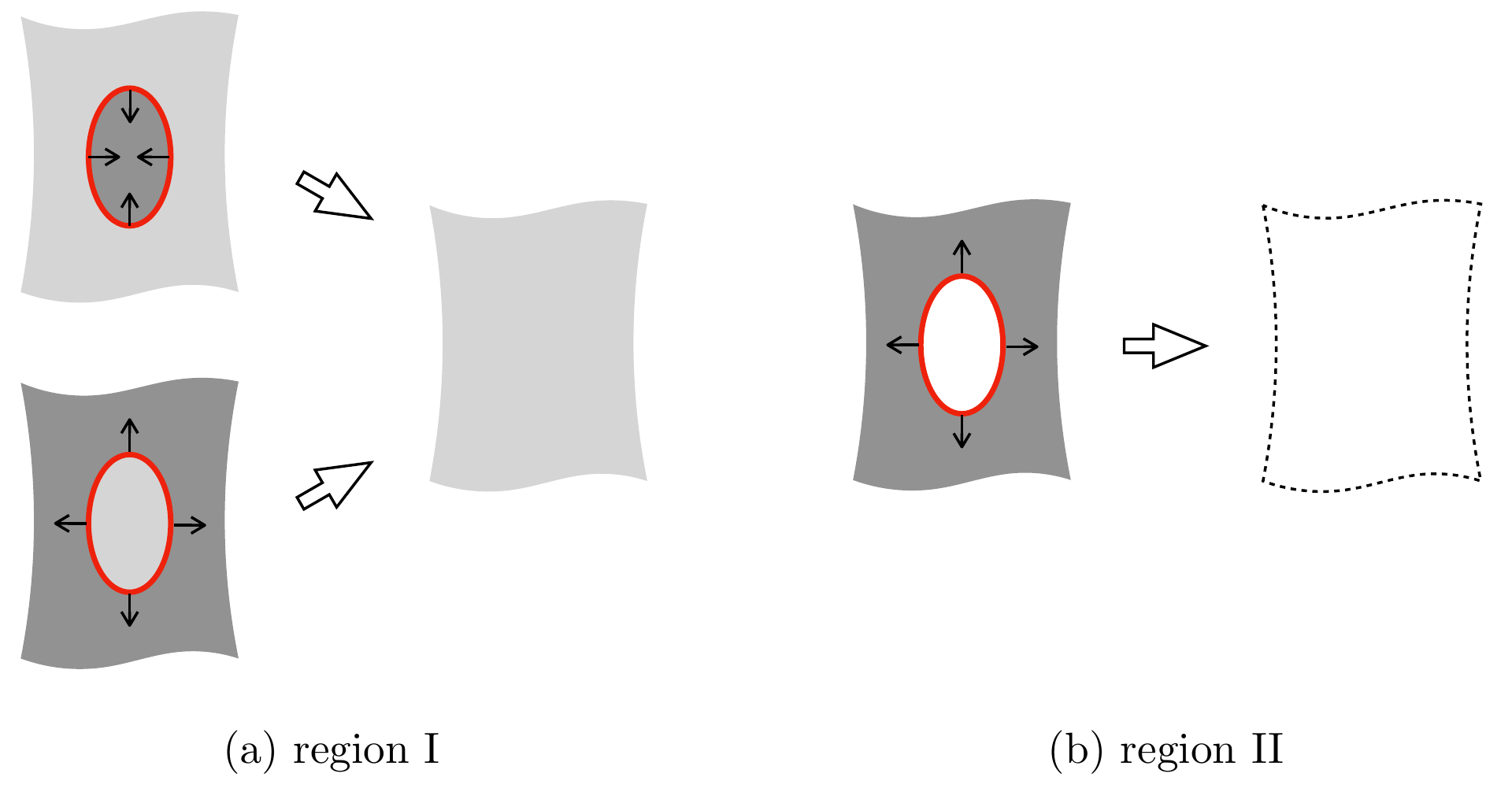}
\caption{(a) A lighter domain wall (less dark) survives creating a hole bounded by the string in the region I. 
(b) A domain wall decays by creating a hole bounded by a string in the region II.}
\label{fig:decay}
\end{center}
\end{figure}

The existence of stable domain walls is problematic since once they are created through the Kibble-Zurek mechanism~\cite{Kibble:1976sj,Zurek:1985qw,Vachaspati-book}, they eventually overclose the Universe. In Region~I, where the vortex is associated with two domain walls attached to it, those domain walls are stable (See left panel of Fig.~\ref{fig:decay} for a schematic picture), therefore cosmologically disfavored. On the other hand, in Region~II, since domain walls have boundaries that are made of vortices, domain walls may decay by creating holes bounded by a vortex loop \cite{Preskill:1992ck,Vilenkin:2000jqa}
(Fig.~\ref{fig:decay} right), therefore it is cosmologically safe. %
In Fig.~\ref{fig:constraint}, parameter space that corresponds to the Region~I is the shaded area (indicated by ``Case II") and the thick-solid line (indicated by ``Case I") and thick-dashed line (indicated by ``Case V"). The parameter space that corresponds to the Region~II is unshaded region in the figure. From the discussion above, the parameter space that is allowed by cosmological consideration is the unshaded region. 
This constraint is quite universal in the sense that it depends only on two parameters in the Higgs sector. It also does not depend on the type of fermion couplings to two Higgs fields at tree level. This is contrasted to existing phenomenological constraints on these parameters, which highly depend on details of model set-up.

Let us briefly discuss here the effect of the explicit CP violation. So far we have assumed that $m_{12}^2$ and $\beta_5$ are real values to make the Lagrangian invariant under the CP transformation shown in Eq.~(\ref{eq:CP}). However, these two parameters can be in general complex. (Only one of them can be made real by a field redefinition.) As a matter of fact, since the fermion sector breaks the CP symmetry, even if we take $m_{12}^2$ and $\beta_5$ to be real values, fermion loop contributions renormalize these parameters to be complex numbers. 
If we allow $m_{12}^2$ and $\beta_5$ to take general complex values, the last line of the Eq.~(\ref{eq:potentialU1a}) becomes $\{-\sin\xi\cos(2\alpha+\delta_1) + \cos\xi\cos(4\alpha+\delta_2)\}$. Here, $\delta_1$ and $\delta_2$ are the complex phase of $m_{12}^2$ and $\beta_5$: $m_{12}^2 = e^{i\delta_1} \vert m_{12}^2 \vert$, $\beta_5 = e^{i\delta_2} \vert \beta_5 \vert$. For non-zero values of $\delta_1$ and/or $\delta_2$, there is always only one global minimum except for the special cases of $\xi=0$ and $\pi$. Therefore the wall-like objects in 2HDMs are in general unstable. Whether those unstable wall-like objects cause cosmological problem or not is determined by the lifetime of the objects. We expect that the smaller the amount of the explicit CP violation is, the longer the lifetime of the wall-like object is. Since the amount of the CP violation is constrained by various phenomenological requirements, such as electric dipole moment measurements (see, e.g., \cite{Keus:2015hva} and references therein), it is interesting to ask whether we can exclude a part of parameter space of 2HDMs by combining cosmological requirement with those phenomenological constraints. 
For such a study, quantitatively reliable estimate of the lifetime of wall-like objects is necessary. Concrete numerical solutions of wall-vortex system that were obtained in this paper is useful for such estimate, which will be done elsewhere.

\section{Conclusion and outlook}
\label{sec:summary}
Two Higgs doublet model is well motivated and the most popular extension of the Higgs sector of the SM. It has four additional scalar degree of freedom which could be detected at future  collider experiments. Masses of extra scalar bosons are determined from model parameters, and existing constraints obtained from various phenomenology, such as bound from direct detection experiment at the LHC and precision EW measurements, are in general highly model dependent. In this paper, we showed that we can place a constraint on a  certain part of the parameter space in 2HDM from the cosmological requirement that stable domain wall should not exist. The constraint obtained in this paper is quite universal in the sense that it does not depend on a choice of parameters other than $m_{12}^2$ and $\beta_5$. It also does not depend on  how SM fermions couple to two Higgs fields at tree level.

Numerical solutions of field configurations of domain wall-vortex system have been obtained, which provide a basis for further quantitative study of cosmology involving such topological objects. These solutions will be also important when we consider domain walls in 2HDM as a localized source of CP violation that could explain the baryon asymmetry of the Universe.

In this paper we have also shown that if one considers the model in the allowed region of the parameter space, domain walls were formed at a certain stage of the early Universe, and should have decayed by creating holes bounded by vortices. The remnant of this formation and decay of domain walls could be detected as a specific spectrum of the gravitational waves~\cite{Saikawa:2017hiv}, which will be studied elsewhere.


\section*{Acknowledgements} 
We would like to thank Motoi Endo, Tomohiro Abe, Kentarou Mawatari, Kazunori Nakayama, Koichi Hamaguchi and Takeo Moroi for useful discussions.
This work is supported by the Ministry of Education, Culture, Sports, Science (MEXT)-Supported Program for the Strategic Research Foundation at Private Universities ``Topological Science'' (Grant No.~S1511006).
The work of M.~E. is supported by KAKENHI Grant Numbers 
26800119, 16H03984 and 17H06462.
The work of M.~K is also supported by MEXT KAKENHI Grant-in-Aid for Scientific Research on Innovative Areas (No.~25105011) and JSPS Grant-in-Aid for Scientific Research (KAKENHI Grant No.~18K03655).
The work of M.~N. is also supported in part by 
JSPS Grant-in-Aid for Scientific Research (KAKENHI Grant No.~16H03984 and No.~18H01217), 
and by a Grant-in-Aid for
Scientific Research on Innovative Areas ``Topological Materials
Science'' (KAKENHI Grant No.~15H05855) 
from the the MEXT of Japan.

\end{document}